\documentclass[12pt]{article}
\usepackage[dvips]{graphicx}
\usepackage{epsfig}
\usepackage{amsmath,amssymb,amsbsy}
\usepackage{amsfonts}
\begin{document}
\thispagestyle{empty}
\begin{center}

{\Large\bf{Determination of the forward slope in $p~p$ and $\bar p~p$
elastic scattering up to LHC energy}}

\vskip1.4cm
{\bf Claude Bourrely}
\vskip 0.3cm
D\'epartement de Physique, Facult\'e des Sciences de Luminy,\\
Universit\'e de la M\'editerran\'ee - Aix-Marseille II,\\
13288 Marseille, Cedex 09, France\\
\vskip 0.5cm
{\bf Jacques Soffer}
\vskip 0.3cm
Physics Department, Temple University\\
Barton Hall, 1900 N, 13th Street\\
Philadelphia, PA 19122-6082, USA
\vskip 0.5cm
{\bf Tai Tsun Wu}
\vskip 0.3cm
Harvard University, Cambridge, MA 02138, USA and\\
Theoretical Physics Division, CERN, 1211 Geneva 23, Switzerland\\
\vskip 0.5cm
{\bf Abstract}\end{center}
In the analysis of experimental data on $p p$ (or $\bar p p$)
elastic differential cross section
it is customary to define an average forward slope $b$ in the form $\exp{(-b|t|)}$, where
$t$ is the momentum transfer. Taking as working example the results of
experiments at Tevatron and SPS, we will show with the help of the impact picture
approach, that this simplifying assumption hides interesting information 
on the complex non-flip scattering amplitude, and that the slope $b$ is not a 
constant.
We investigate the variation of this slope parameter, including a model-independent 
way to extract this information from an accurate measurement of the elastic
differential cross section.
An extension of our results to the LHC energy domain is presented in view of future 
experiments.

\vskip 0.5cm

\newpage
%%%%%%%%%%%%%%%%%%%%%%%%%%%%%%%%%%%%%%%%%%%%%%%%%%%%%%%%%%%%%%%%%%%%%%
\section{Introduction}
High energy  $\bar p p~\mbox{and}~p p$ elastic scattering measured at ISR, SPS,
and Tevatron colliders
have provided usefull informations on the behavior of the scattering
amplitude, in particular, on the nature of the Pomeron. A large step in energy
domain is accomplished with the LHC collider presently running, giving a unique
opportunity to improve our knowledge on the asymptotic regime of the scattering
amplitude.
The measurement of the differential cross section in previous experiments
has shown the existence of a dip and a shrinkage of the diffraction peak with
increasing energy. One information deduced from this measurement concerns
the forward slope. It is generally assumed that the cross section near the
forward direction behaves like $\exp{(-b|t|)}$, where the slope $b$ is treated on 
the average as a constant, which increases with the energy. 
In fact, from theoretical models, one can compute the slope with the result 
that it has an interesting behavior as a function of $t$, linked to the analytic
structure of the complex scattering amplitude. This feature was already noticed
in Fig. 5 of Ref. \cite{bou9}.

In section \ref{sec1} we will show using the impact 
picture approach (BSW)
\cite{bou9}-\cite{bou8} developped many years ago, which has proven to give
a reliable predictions of hadron-hadron elastic scattering, that the behavior
of the complex scattering amplitude has an impact on the behavior of the
forward slope. We will focus on SPS and Tevatron experiments, 
which give at the moment the highest available energy data, whereas ISR data 
 could have been used as well. 
In section \ref{sec2} we discuss the possibility to extract information on the 
slope from a measurement of the differential cross, without 
making reference to any theoretical model.
In section \ref{sec3} we extend our analysis to the LHC energy
domain where we present some predictions in view of planed experiments by 
ATLAS/ALFA and TOTEM collaborations.

\section{The slope behavior in the forward direction}
\label{sec1}
\begin{figure}[ht]
 \vspace*{-15mm}
\begin{center}
  \begin{minipage}{6.5cm}
  \epsfig{figure=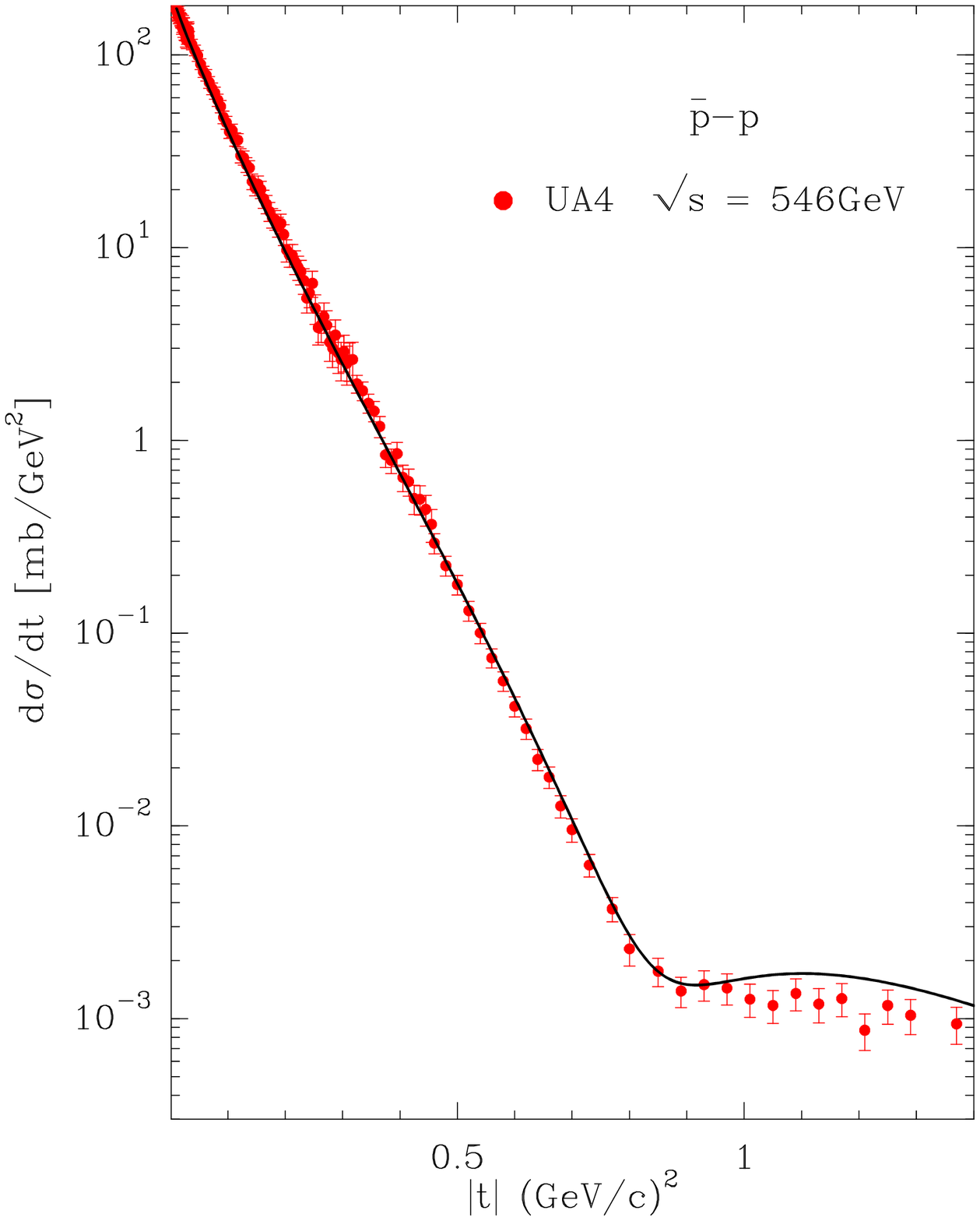,width=8.0cm}
  \end{minipage}
    \begin{minipage}{6.5cm}
  \epsfig{figure=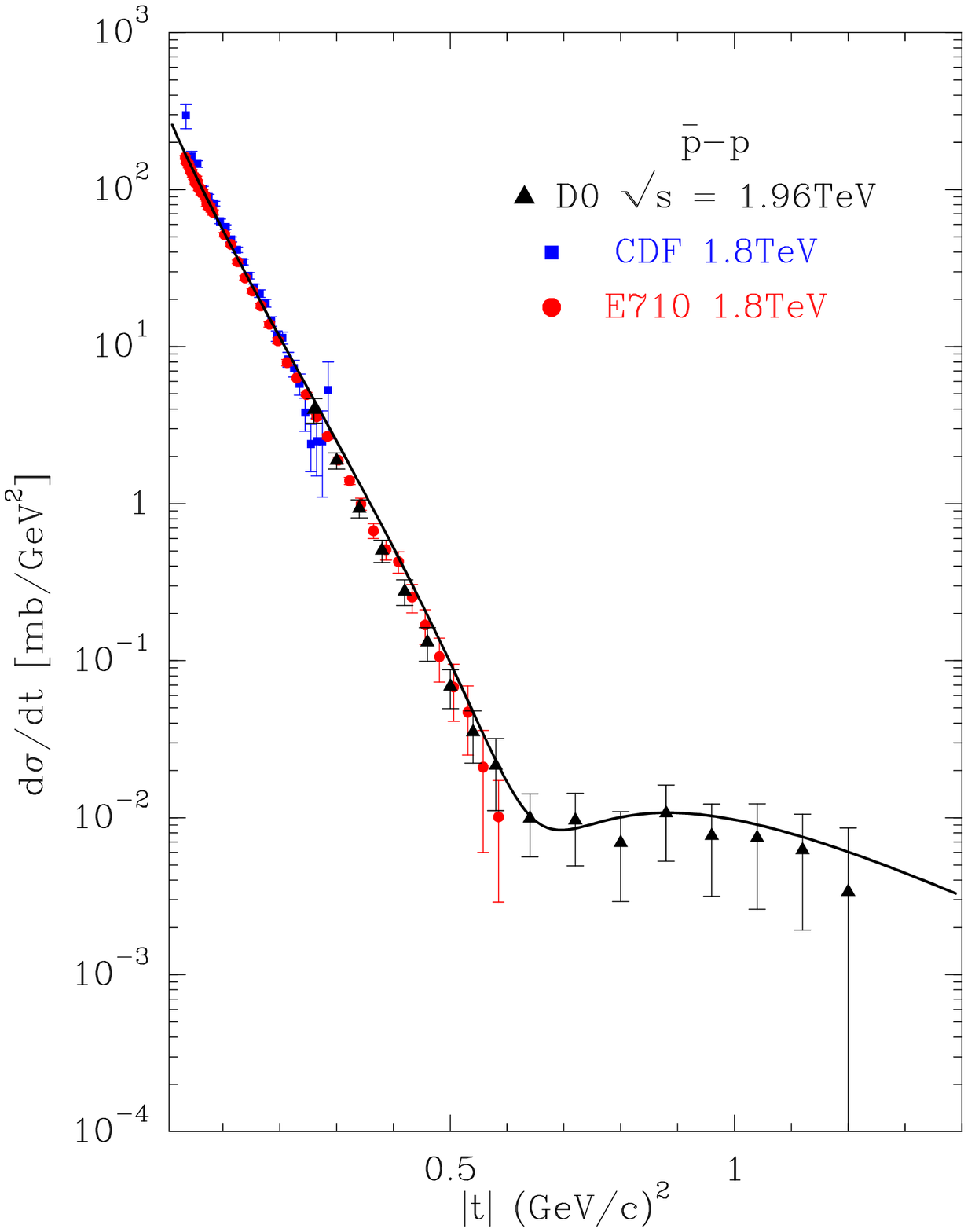,width=8.0cm}
    \end{minipage}
\end{center}
  \vspace*{-15mm}
\caption{The $\bar p p$ elastic differential cross section
 as a function of $t$, predicted by the BSW approach (solid curves).
Data from UA4 collaboration  \cite{ua43} at $\sqrt{s} = 546 \mbox{GeV}$ (left).
The $\bar p p$ elastic differential cross section at 
$\sqrt{s} = 1.8-1.96 \mbox{TeV}$. Experimental data
are from D0  \cite{D0} preliminary (triangle), CDF \cite{cdf} (square), 
E710  \cite{e710} (circle) collaborations
(right).}
\label{dsigcross}
\vspace*{-1.5ex}
\end{figure}

\begin{figure}[htp]
 \vspace*{-15mm}
\begin{center}
  \begin{minipage}{6.5cm}
  \epsfig{figure=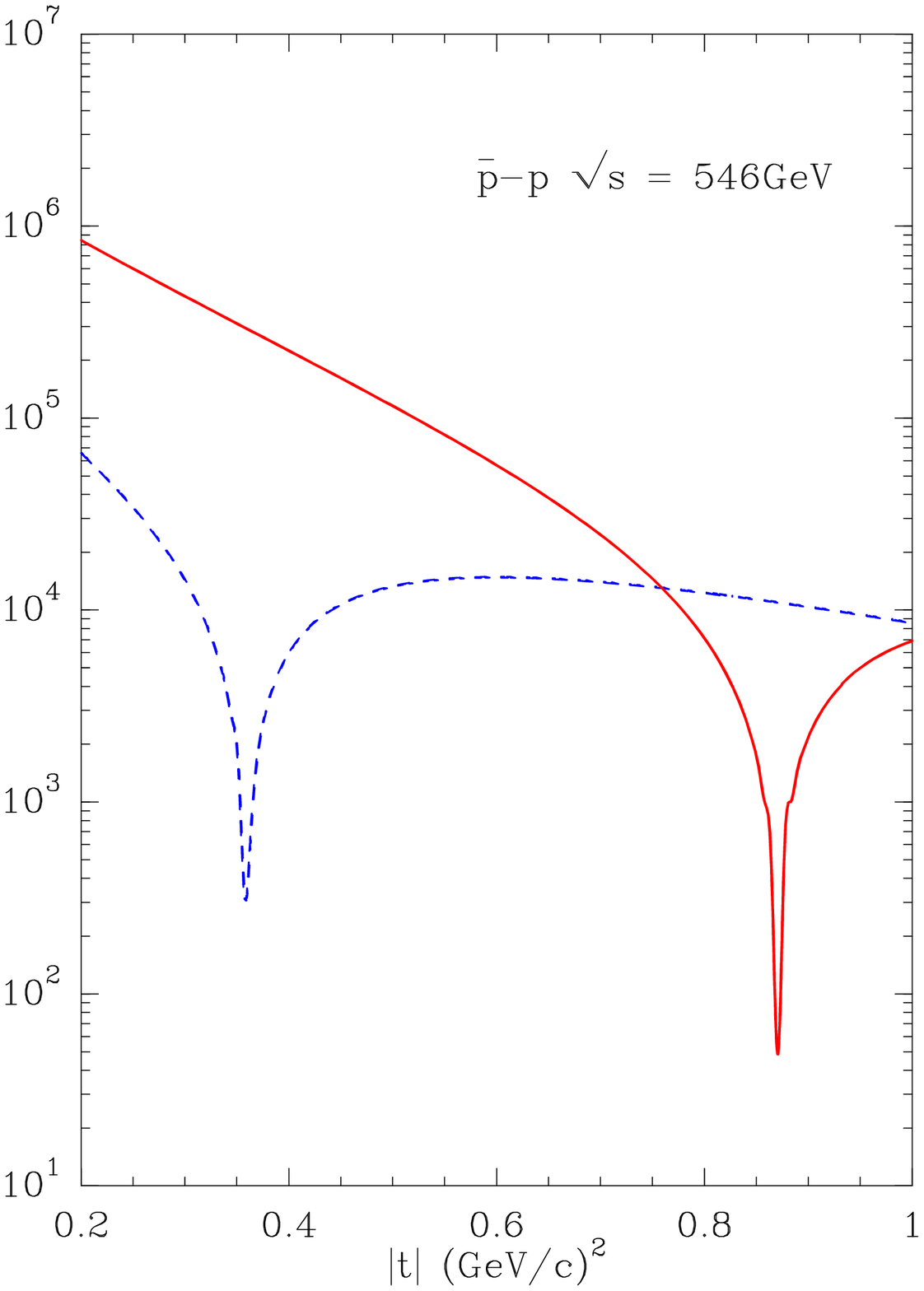,width=8.0cm}
  \end{minipage}
    \begin{minipage}{6.5cm}
  \epsfig{figure=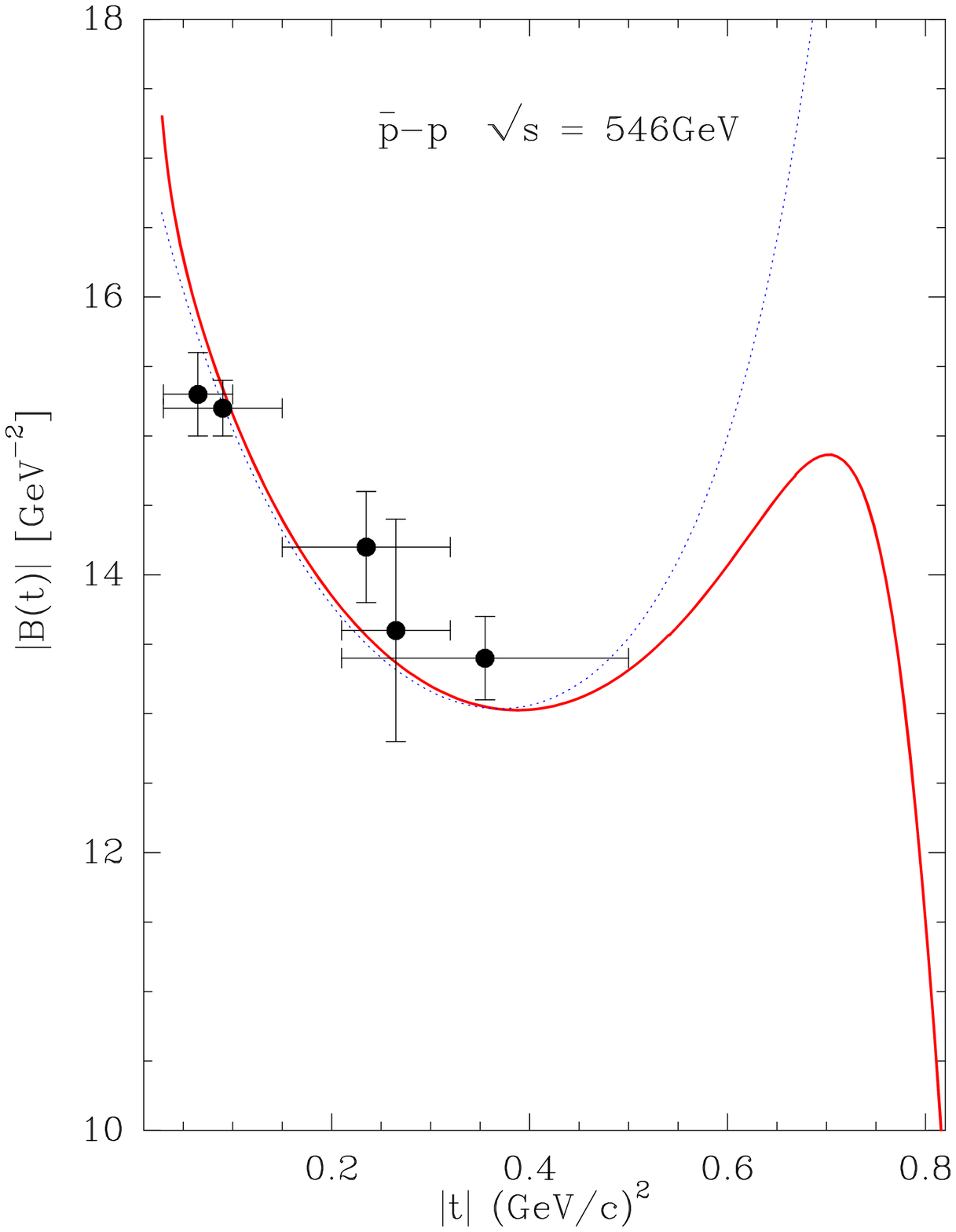,width=8.0cm}
    \end{minipage}
\end{center}
  \vspace*{-10mm}
\caption{The absolute value of the $\bar p p$ scattering amplitude, 
solid curve $|$Imag$|$, dashed curve
$|$Real$|$ as a function of $t$ for $\sqrt{s}=$ 546GeV (left).
The forward slope as a function of $t$ calculated from the impact picture
approach. $\bar p p$ scattering solid curve, the dotted curve 
corresponds to the real part of the amplitude set to zero.
Experimental data are from UA4 collaboration \cite{ua42} (right).}
\label{slopeua4}
\vspace*{-1.5ex}
\end{figure}

In the impact picture approach \cite{bou9} we define the scattering amplitude as 
\footnote{The details of the parameters and formulas
are given in \cite{bou9}-\cite{bou8}}
\begin{equation}\label{amplibsw}
a(s,t) = \frac{is}{2\pi}\int e^{-i\mathbf{q}\cdot\mathbf{b}} (1 - 
e^{-\Omega_0(s,\mathbf{b})})  d\mathbf{b} \ ,
\end{equation}
where $\mathbf q$ is the momentum transfer ($t={-\bf q}^2$) and 
$\Omega_0(s,\mathbf{b})$ is the opaqueness at impact parameter 
$\mathbf b$ and at a given energy $s$. We take 
\begin{equation}\label{opac}
\Omega_0(s,\mathbf{b}) = S_0(s)F(\mathbf{b}^2)+ R_0(s,\mathbf{b}) \, ,
\end{equation}
the first term is associated with the "Pomeron" exchange, which generates 
the diffractive component of the scattering and the second term is 
the Regge background which is negligible at high energy.
The Pomeron energy dependence is given by the complex
crossing symmetric expression 
\begin{equation}\label{energ}
S_0(s) = {s^c \over (\ln s)^{c'}} + {u^c \over (\ln u)^{c'}} \, ,
\end{equation}
where $u$ is the third Mandelstam variable.
This implies that the Pomeron is a fixed Regge cut rather than a
Regge pole.
The choice one makes for $F(\mathbf{b}^2)$ is essential
and we take the Bessel 
transform of
\begin{equation}\label{formf}
\tilde F(t) = f[G(t)]^2 {a^2 + t \over a^2 -t} \, ,
\end{equation}
where $G(t)$ stands for the proton "nuclear form factor", parametrized like 
the electromagnetic form factor, as having two poles, 
\begin{equation}\label{fgt}
G(t) = {1 \over (1 - t/m_1^2)(1 - t/m_2^2)} \, .  
\end{equation}
The total cross section reads
\begin{equation}\label{sigtot}
\sigma_{tot} (s) = \frac{4\pi}{s} \mbox{Im}~a(s, t=0) \,, 
\end{equation}
and the differential cross section
\begin{equation}\label{dsigdt}
{d\sigma (s,t) \over dt} = \frac{\pi}{s^2}|a(s,t)|^2 \,.
\end{equation}
The slope of the differential cross section is given by
\begin{equation}
B(t) = \frac{d}{dt}\log{(\frac{d\sigma}{dt})}~.
\label{eq1}
\end{equation}
In order to have a global view of our predictions over the 
full measured $t$ range, the elastic
differential cross sections are plotted in Fig. \ref{dsigcross} for the
UA4 experiment at $\sqrt{s} = 546\mbox{GeV}$ \cite{ua41}-\cite{ua43}
and for the D0, CDF, E710 experiments at $\sqrt{s} = 1.8-1.96\mbox{TeV}$
\cite{e710}-\cite{D0}. A comparison of the prediction with experimental data
 gives respectively a $\chi^2$/pt = 1.12  for UA4 and 
1.2 for the Tevatron energies. In order to show the relation between the
scattering amplitude and the slope near the forward direction, we plot
in Fig. \ref{slopeua4} the absolute value of the real and imaginary
parts of the $\bar p p$ amplitude at $\sqrt{s} = 546\mbox{GeV}$ 
together with the forward slope as a function of $t$. One observes that the real
part of the amplitude is negligible below $|t| = 0.6\mbox{GeV}^2$ or so and clearly
around $|t| = 0.8\mbox{GeV}^2$, its contribution fills up the dip of the cross
section, as seen in Fig. \ref{dsigcross}.
%\clearpage
%\newpage
 Concerning the behavior of the slope, 
the impact picture predicts a minimum around $|t| = 0.35\mbox{GeV}^2$ and this 
is in agreement with the slopes obtained by the UA4 experiment \cite{ua42}, 
in different $t$ intervals. 
The dotted curve corresponds to the variation of the slope obtained when
the real part of the amplitude is set to zero and,  as expected, it shoots up 
near the zero of the imaginary part.
\begin{figure}[htbp]
  \vspace*{-15mm}
\begin{center}
  \begin{minipage}{6.5cm}
  \epsfig{figure=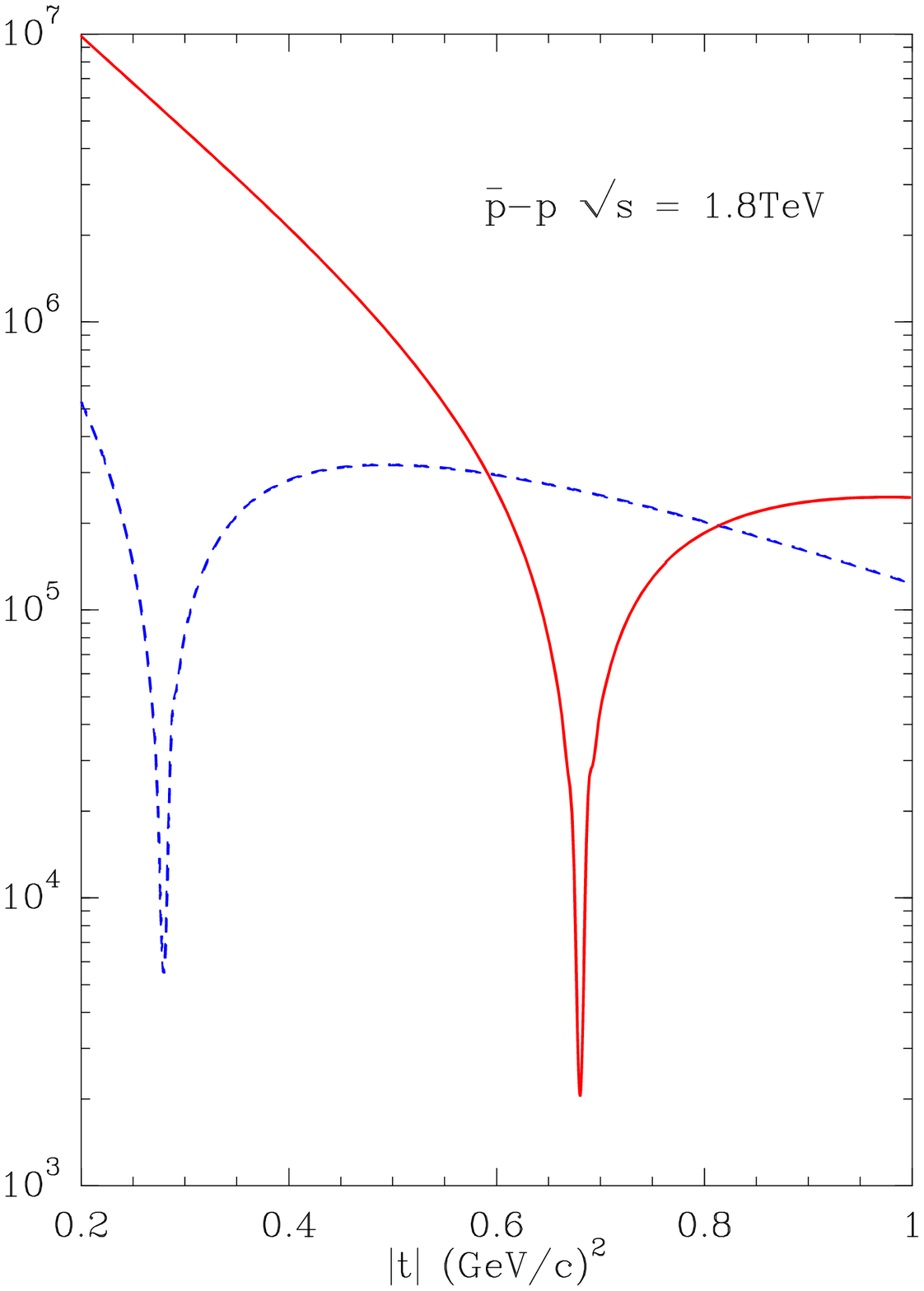,width=8.0cm}
  \end{minipage}
    \begin{minipage}{6.5cm}
  \epsfig{figure=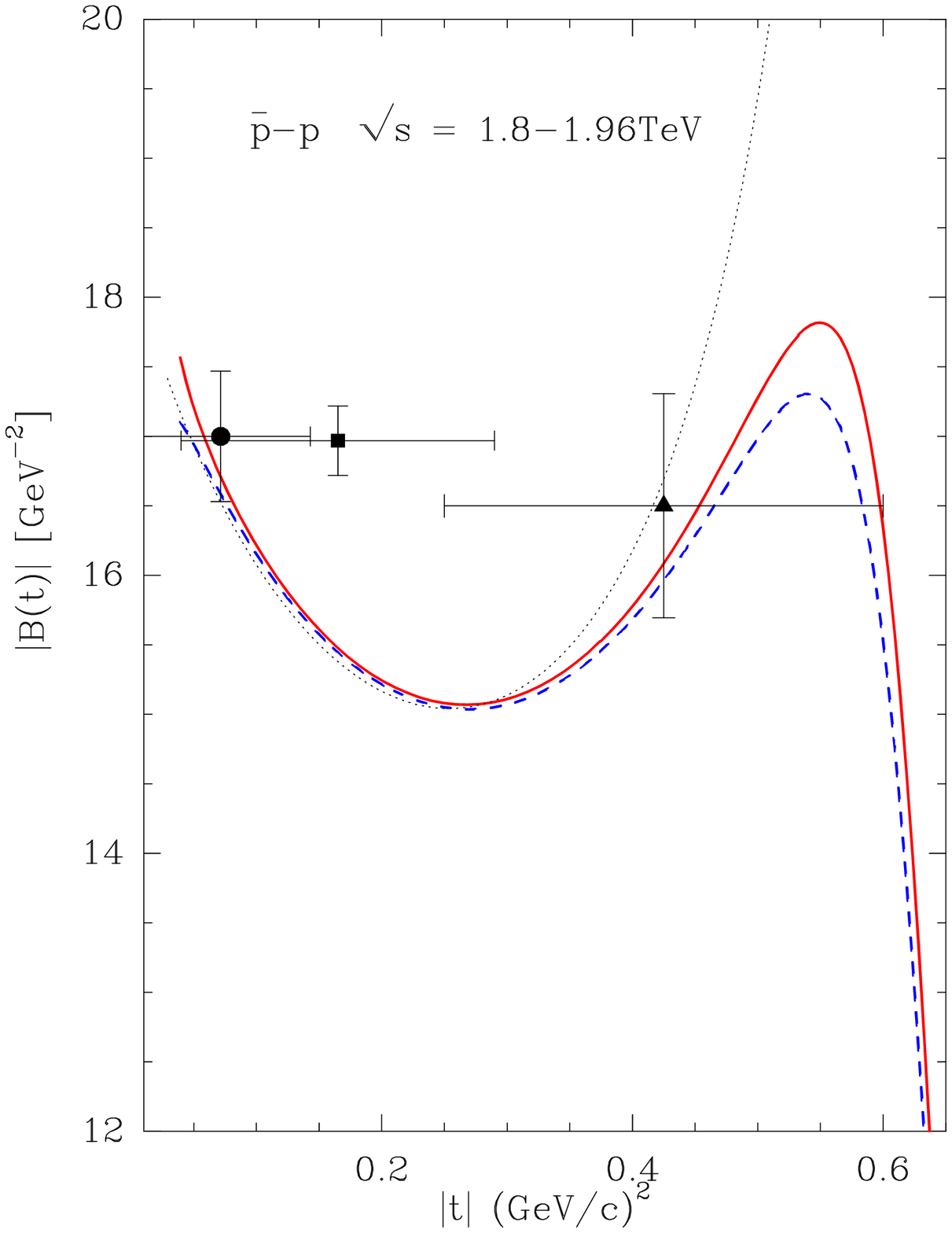,width=8.0cm}
    \end{minipage}
\end{center}
  \vspace*{-15mm}
\caption{The absolute value of the $\bar p p$ scattering amplitude,
 solid curve $|$Imag$|$, dashed curve
$|$Real$|$ as a function of $t$ for $\sqrt{s}=$ 1.8TeV (left).
The forward slope as a function of $t$ calculated from the impact picture 
approach. $\bar p p$ scattering solid curve, $p p$ dashed curve, 
dotted curve the real part of the amplitude set to zero.
Experimental data are from D0  preliminary (triangle), CDF (square), 
E710 (circle) collaborations (right).}
\label{slopetevat}
\vspace*{-1.5ex}
\end{figure}
At Tevatron energies we show in Fig. \ref{slopetevat} the $\bar p p$ 
absolute value of real and imaginary parts of the elastic 
scattering amplitude at $\sqrt{s} = 1.8\mbox{TeV}$. We observe that the zero of the
real and imaginary parts 
have moved toward smaller $t$ values compared to the SPS energy, an effect
which was predicted theoretically a long time ago and is
responsible for the shrinkage of the diffraction peak. The behavior of the
slope (solid curve) shows again a minimum around $|t| = 0.25\mbox{GeV}^2$.
A comparison with the experimental slopes measured by D$\varnothing$ \cite{D0},
CDF \cite{cdf} and E710 \cite{e710} agrees with our prediction, with no clear
indication of a mininum, although the errors are large. 
At high energy the differential cross sections for $p p$
and $\bar p p$ become close, the dashed curve represents the slope for the
$p p$ case, and here also we have plotted with a dotted curve the slope when
the real part of the amplitude is set to zero.

\section{Extraction of the slope from experimental measurements}
\label{sec2}
\begin{figure}[htbp]
  \vspace*{-15mm}
\begin{center}
  \begin{minipage}{6.5cm}
  \epsfig{figure=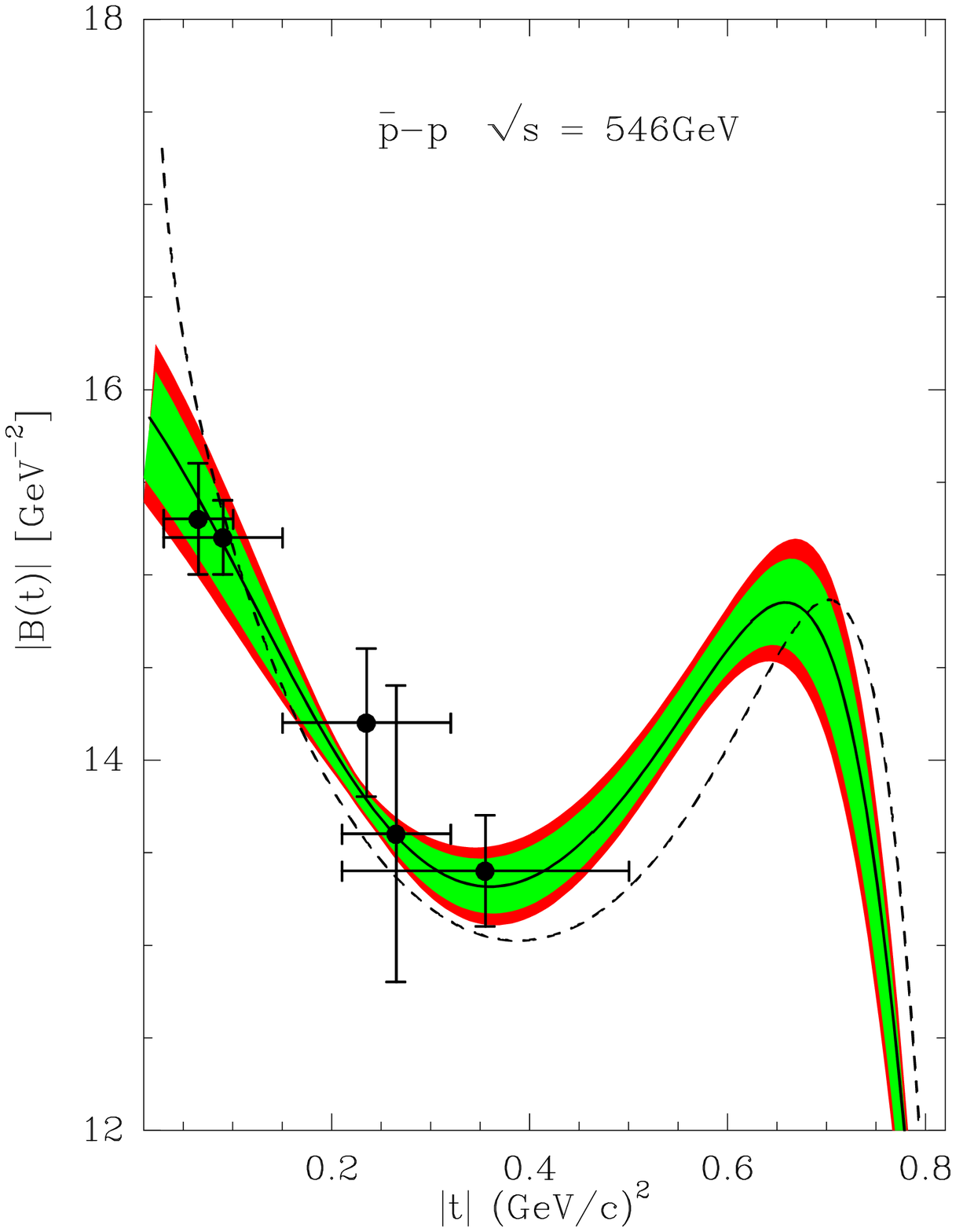,width=8.0cm}
  \end{minipage}
    \begin{minipage}{6.5cm}
  \epsfig{figure=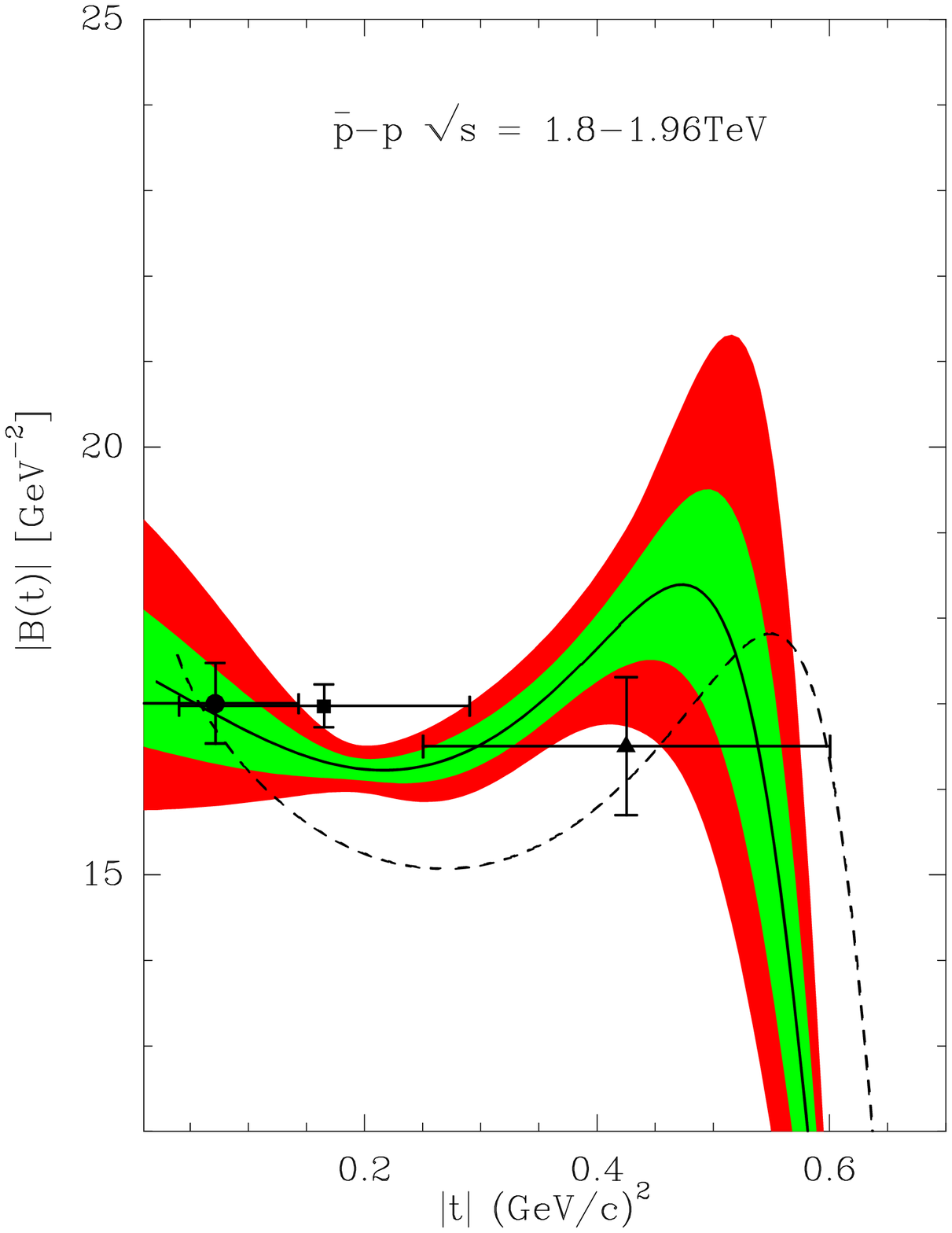,width=8.0cm}
    \end{minipage}
\end{center}
  \vspace*{-10mm}
\caption{The slope $B(t)$ as a function of $t$ obtained from experiment,
left  $\sqrt{s}=$ 546GeV solid curve,  
right $\sqrt{s}=$ 1.8TeV solid curve, BSW prediction dashed curve. 
Inner bands represent the uncertainties with a CL 68\% and outer bands 95\%.
Same experimental data as in Figs.
\ref{slopeua4}, \ref{slopetevat}.}
\label{slopexp}
\vspace*{-1.5ex}
\end{figure}

In the previous section we have made an analysis by a combination of 
theoretical
and experimental results, but we would like to examine the situation when only an
experimental information is available, i.e., a model independent analysis.
We will use the examples of the previous section.
Let us make a preliminary remark: it is customary to determine the forward
slope by a fit of the cross section containing an expression like $\exp{(bt)}$ or
$\exp{(bt +ct^2)}$. It turns out that this approximation is only valid in a very 
narrow range of $t$ because as we have seen in the previous section, the
slope $b$ cannot be treated as a constant over a $t$ domain up to the
first minimum of the cross section \footnote {A simple multiplication of the exponential 
by a polynomial in $t$, or a Spline function, induces unwanted oscillations.}.
So we propose to define a numerical parametrization of the scattering amplitude 
in a limited $t$ range ($t \le 0$) taking into account the analytic properties
of the amplitude as follows:
\begin{eqnarray}
\mbox{Re}~a_p(t) &=& c_1(t_0 + t)e^{d_1 t} \,,  \label{rap} \\
\mbox{Im}~a_p(t) &=& c_2(t_1 + t)e^{d_2 t} \,,  \label{iap} \\
\frac{d\sigma(t)}{dt} &=& (Re~a_p(t))^2 + (Im~a_p(t))^2 \,,\label{dsigp}
\end{eqnarray}
and the slope is given by
\begin{equation}
B(t) = \frac{2c_1^2 (t_0 +t)(1 +d_1(t_0 +t))e^{2 d_1 t}+
2c_2^2 (t_1 +t)(1 +d_2(t_1 +t))e^{2 d_2 t}}{d\sigma(t)/dt} \,.
\label{slopep}
\end{equation}
For UA4 making a fit of the differential cross section in the range 
$ 0.1 \leq -t \leq 0.8$  we get for the parameters of Eq. (\ref{rap}-\ref{iap})
the values
\begin{equation}\label{num1}
 c_1 = 30.7\pm 1.9 \quad c_2 =  -12.4 \pm 0.98 \quad d_1 = 7.06\pm 0.18 
\quad d_2 =  4.58 \pm 0.2\,,
\end{equation}
\begin{equation}\label{num2}
t_0 = 0.33\pm 0.04 \quad t_1 = 0.81 \pm 0.007\,,
\end{equation}
with a $\chi^2$/pt = 1.2, where the units are $\mbox{GeV}^2$, for $t$, $t_0$, $t_1$,
 $\mbox{GeV}^{-2}$, for $d_1$, $d_2$ and $\sqrt{\mbox{mb}}$/\mbox{GeV}, for $c_1$ and $c_2$.\\
The high accuracy of the UA4
experiment allows a determination of the slope variation (solid line, 
Fig. \ref{slopexp} left) in perfect agreement with the experimental values 
obtained in \cite{ua42}. We show the uncertainties which correspond to a confidence
level (CL) 68\% inner bands and 95\% outer bands.
We also show for comparison  the BSW prediction (dashed line). The values
of the parameters $t_0,~t_1$ in (\ref{num2}) are close to the  zeros obtained
from the BSW approach (see Fig. \ref{slopeua4}). Since we are working 
with a numerical parametrization the values of the other parameters in 
(\ref{num1}) have no particular meaning.
We have made the same type of analysis
for Tevatron energies where we obtain from a fit of the experimental
data from E710 \cite{e710}, CDF \cite{cdf}, D$\varnothing$ \cite{D0},
in the range $ 0.095 \leq -t \leq 0.96$ the following values:
\begin{equation}\label{num3}
c_1 = 42.41 \pm 9.2 \quad c_2 =  -23.14 \pm 4.9 \quad d_1 = 7.82 \pm 0.85
 \quad d_2 =  5.06 \pm 0.5\,,
\end{equation}
\begin{equation}\label{num4}
 t_0 = 0.28\pm 0.1 \quad t_1 = 0.6\pm 0.02\,,
\end{equation}
with a $\chi^2$/pt = 0.75, it seems that we get a better fit compared to UA4, but
 this is due to larger errors. With these parameters the slope variation
is shown in Fig. \ref{slopexp} right (solid line) with the corresponding
experimental measurements, they are both compatible within the experimental
errors, however some comments are in order. 
\begin{table}[htb]
\begin{center}
\begin{tabular}{|c|c|c|}
\hline
&&\\
$\sqrt{s}$ (TeV) & Real  & Imaginary\\
&&\\
\hline\raisebox{0pt}[12pt][6pt]
0.1    & 0.475    & 1.201 
\\[4pt]
\hline\raisebox{0pt}[12pt][6pt]
0.2    & 0.407    & 1.050 
\\[4pt]
\hline\raisebox{0pt}[12pt][6pt]
0.5    & 0.345    & 0.886 
\\[4pt]
\hline\raisebox{0pt}[12pt][6pt]
1.0    & 0.297    & 0.776 
\\[4pt]
\hline\raisebox{0pt}[12pt][6pt]
1.8    & 0.256    & 0.667 
\\[4pt]
\hline\raisebox{0pt}[12pt][6pt]
5.0    & 0.202    & 0.544 
\\[4pt]
\hline\raisebox{0pt}[12pt][6pt]
7.0    & 0.193    & 0.509 
\\[4pt]
\hline\raisebox{0pt}[12pt][6pt]
10.0    & 0.180    & 0.473 
\\[4pt]
\hline\raisebox{0pt}[12pt][6pt]
14.0    & 0.168    & 0.442 
\\[4pt]
\hline\raisebox{0pt}[12pt][6pt]
20.0    & 0.156    & 0.412 
\\[4pt]
\hline
\end{tabular}
\caption {Zeros of the real and imaginary nuclear amplitude  
for $p~p$ elastic in BSW}
\label{table1}
\end{center}
\vspace*{-1.5ex}
\end{table}
In the determination of the slope
variation we have used the range $ 0.095 \leq -t \leq 0.96$, we notice that
the CDF data are limited to $-t \leq 0.285$, the D$\varnothing$ data start
at $-t \geq 0.26$ and are preliminary, so both experiments do not cover
the required range in $t$. Only E710 has data points inside the domain of
analysis, but as noticed in \cite{cdf}, this measurement seems to be less accurate.
Altogether, we have obtained for the slope a variation less precise than
in the UA4 case, this is also reflected by the discrepancy with the prediction
made with the BSW approach. In Eqs. (\ref{rap}-\ref{iap}) we have introduced
a zero in the amplitudes, in order to have a reference set we give 
in Table \ref{table1}
the values of the zeros obtained in BSW for the real and imaginary of the nuclear
amplitude for $p~p$ elastic as a function of $\sqrt{s}$.

\section{The LHC energy domain}
\label{sec3}
In view of experiments planed at LHC to measure the $p~p$ total cross section and
elastic scattering by TOTEM \cite{totem} and ATLAS/ALFA \cite{alfa}, 
it is of interest to make some predictions on the slope behavior in this energy 
domain.
\begin{figure}[htp]
  \vspace*{-20mm}
\begin{center}
  \epsfig{figure=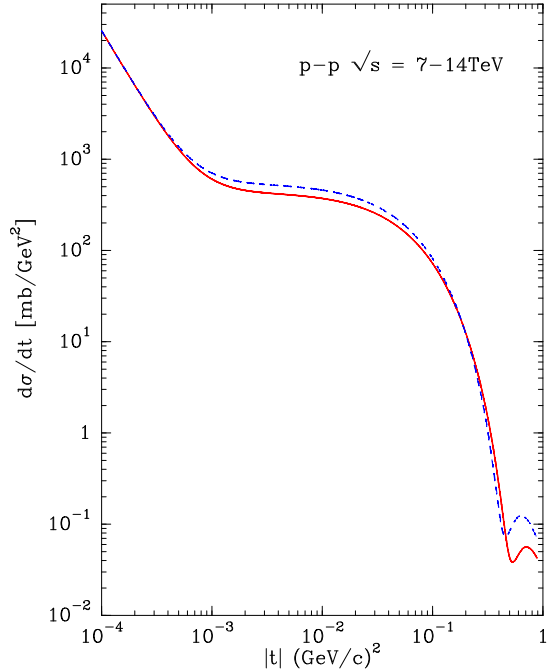,width=8.5cm}
\end{center}
  \vspace*{-15mm}
\caption{
The $p p$ elastic differential cross section as a function of $t$, solid curve
7TeV, dashed curve 14TeV as predicted by the BSW approach.}
\label{dsiglhc}
\vspace*{-1.5ex}
\end{figure}
The differential cross sections for LHC energies
$\sqrt{s} =$ 7-14TeV  as a function of $t$
in a log-log scale are plotted  in Fig. \ref{dsiglhc}. We observe
a continuous change of curvature above the Coulomb region up to the first
minimum.
Looking at the behavior of real and imaginary parts of the full amplitude
at low $t$, including the nuclear and Coulomb contributions,
we notice that the real part has two zeros, one located at
$-t = 0.0064$ which comes from a destructive interference between the Coulomb and
the nuclear part and a second one at
$-t = 0.18$ due to the nuclear part alone. 
The nuclear imaginary part has one zero located at $-t = 0.5$ 
(see Fig. \ref{slopelhc} left). Notice that for the reaction $\bar p~p$ 
the interference is constructive so the first zero
in the real part does not exist.
%\clearpage
\newpage
\begin{figure}[htbp]
  \vspace*{-15mm}
\begin{center}
  \begin{minipage}{6.5cm}
  \epsfig{figure=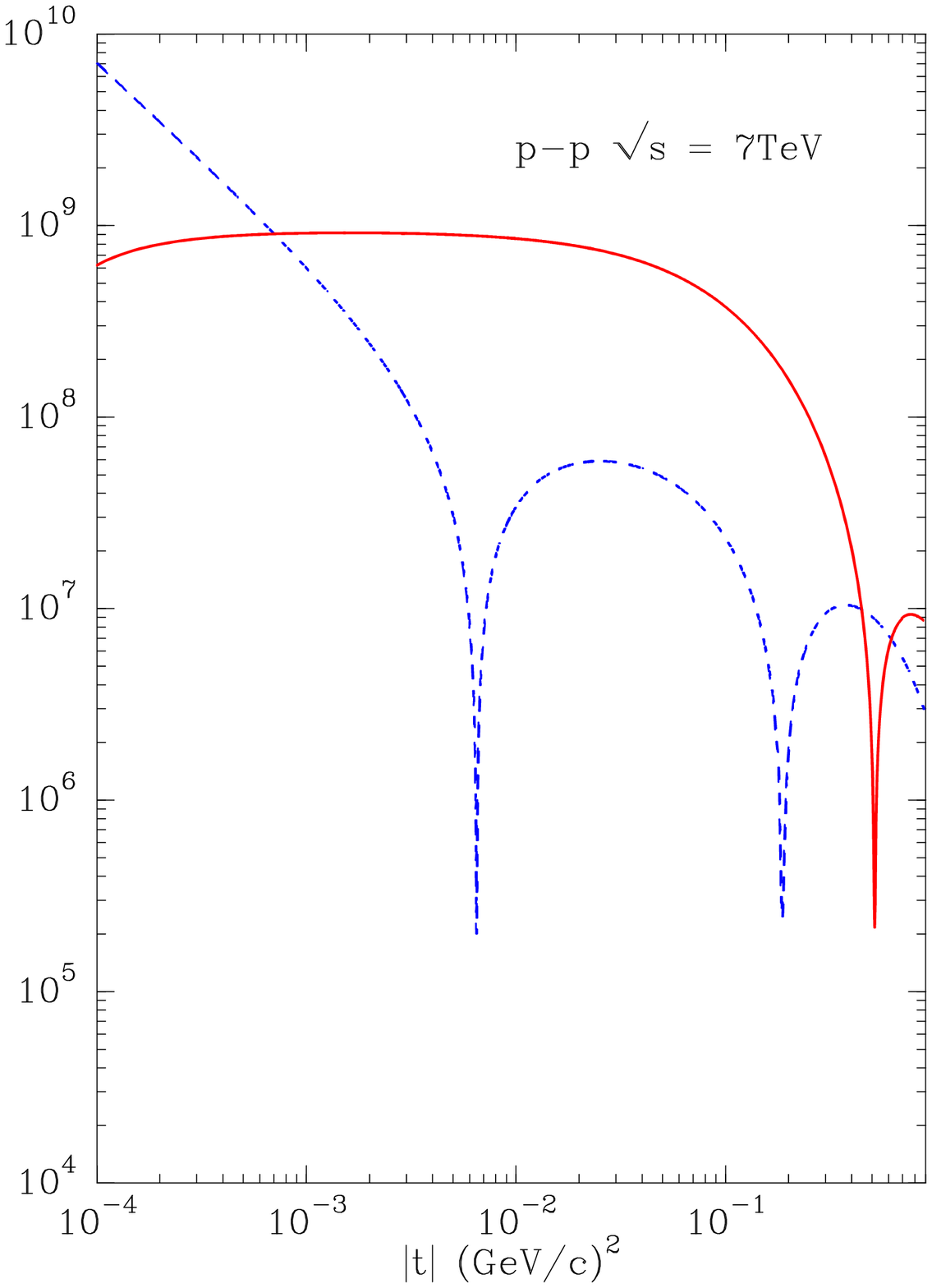,width=8.0cm}
  \end{minipage}
    \begin{minipage}{6.5cm}
  \epsfig{figure=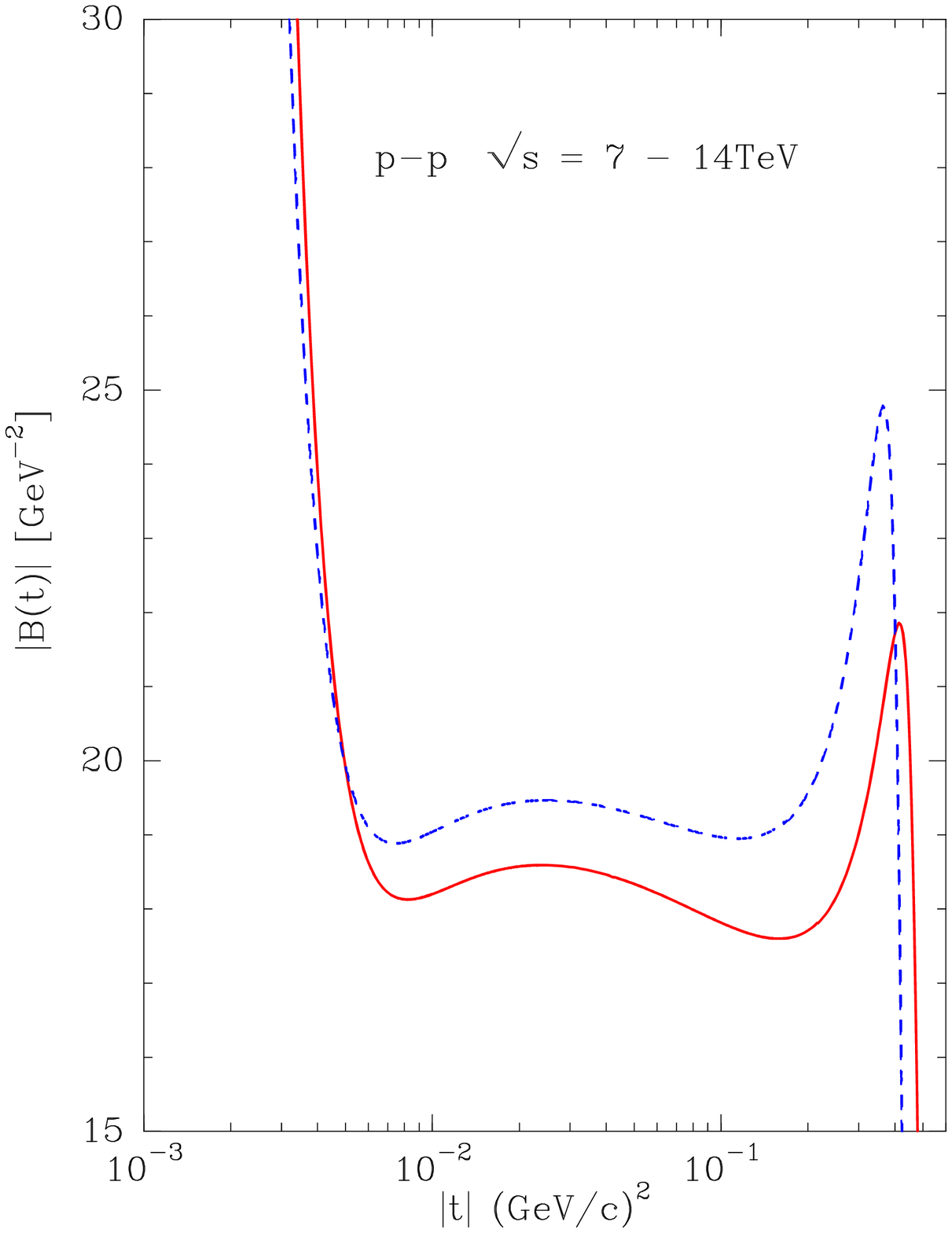,width=8.0cm}
    \end{minipage}
\end{center}
  \vspace*{-15mm}
\caption{The absolute value of the $p p$ elastic scattering 
amplitudes at $\sqrt{s} =$ 7 TeV as a function of $t$ from the BSW approach, 
real part dashed curve, imaginary part solid (left). 
The forward slope as a function of $t$,  7TeV solid curve, 14TeV 
dashed curve (right).}
\label{slopelhc}
\vspace*{-1.5ex}
\end{figure}
The effect of these zeros on the slope variation is illustrated in Fig. 
\ref{slopelhc} right, 
for two energies $\sqrt{s} = 7\mbox{TeV}$ solid curve, and 14TeV dashed curve,
a  minimum occurs for $-t = 0.007$ and $-t = 0.15$.
It is clear that precise mesurements will be required to put in evidence the 
oscillating structure of the slope. At lower energy the RHIC collider has a
project to measure the $p~p$ differential cross at $\sqrt{s} =$ 500GeV, we expect
for the slope a behavior similar to Fig. \ref{slopelhc}.

\section{Concluding remarks}
The measurement of the elastic differential cross section in $p~p$ and 
$\bar p~p$ has revealed
the existence of a dip at large $t$ starting at the ISR energies 
which is due to the dominance
of the real part of the scattering amplitude over the imaginary part, this feature
induces an important change in the slope of the cross section. For low momentum
transfer it is supposed that the slope is constant and a form like
$\exp{(-b|t|)}$ is sufficient to describe the cross section behavior. However,
the computation of the slope with theoretical models shows that this approximation
is too crude and due to the existence of zeros in the real and imaginary parts 
the variation
of the slope has a more complicated structure.

In an impact picture approach (BSW) we have computed the behavior of the slope in
the forward region making evidence of a minimum followed by a maximum corresponding
to the dip region. Although the measured slopes are compatible with our results
taking into account the experimental errors, a more precise analysis of the
data is needed to reveal the structure of the slope connected with the amplitude
behavior. In order to compute the slope from experimental data alone, 
we have proposed a simple numerical parametrization of the amplitude 
that we have applied to the UA4 and Tevatron cross section measurements. 
The slope behavior so obtained reproduces the theorical one, 
but its determination depends strongly on the precision of the experimental data.

In view of the future experiments planed at LHC we have made predictions for
two energies $\sqrt{s} =$ 7 and 14TeV. Our results show for the slope a more
intricate behavior with momentum transfer, which can be put in evidence
if very precise experimental data are obtained.

\end{document}